\documentclass[12pt]{iopart}
\pdfoutput=1
\expandafter\let\csname equation*\endcsname\relax
\expandafter\let\csname endequation*\endcsname\relax

\usepackage{color}
\usepackage[breakwords]{truncate}
\usepackage[font={small}]{caption}
\usepackage{physics}
\usepackage{graphicx}
\usepackage{cite} 
\usepackage{subcaption}
\usepackage{etoolbox}
\usepackage[section]{placeins}
\usepackage[sectionbib]{chapterbib}
\usepackage{amsfonts}
\usepackage[bookmarks,bookmarksopen,bookmarksdepth=2]{hyperref}

\makeatletter
\newrobustcmd{\fixappendix}{%
  \patchcmd{\l@section}{1.5em}{7em}{}{}%
  \patchcmd{\l@subsection}{2.3em}{7em}{}{}%
}
\makeatother

\appto\appendix{
\addtocontents{toc}{\fixappendix}
\addtocontents{toc}{\protect\setcounter{tocdepth}{1}}}

\begin{document}
\newcommand{\ben}[1]{\textcolor{blue}{\textbf{#1}}}

\title{Generating constrained run-and-tumble trajectories}
\author{Benjamin De Bruyne}
\address{LPTMS, CNRS, Univ.\ Paris-Sud, Universit\'e Paris-Saclay, 91405 Orsay, France}
\author{Satya N. Majumdar}
\address{LPTMS, CNRS, Univ.\ Paris-Sud, Universit\'e Paris-Saclay, 91405 Orsay, France}
\author{Gr{\'e}gory Schehr }
\address{Sorbonne Universit\'e, Laboratoire de Physique Th\'eorique et Hautes Energies, CNRS UMR 7589, 4 Place Jussieu, 75252 Paris Cedex 05, France}
\eads{\mailto{benjamin.debruyne@centraliens.net}, \mailto{satya.majumdar@universite-paris-saclay.fr},
\mailto{gregory.schehr@u-psud.fr}}
\begin{abstract}
We propose a method to exactly generate bridge run-and-tumble trajectories that are constrained to start at the origin with a given velocity and to return to the origin after a fixed time with another given velocity. The method extends the concept of effective Langevin equations, valid for Markovian stochastic processes such as Brownian motion, to a non-Markovian stochastic process driven by a telegraphic noise, with exponentially decaying correlations. We obtain effective space-time dependent tumbling rates that implicitly accounts for the bridge constraint. We extend the method to other types of constrained run-and-tumble particles such as excursions and meanders. The method is implemented numerically and is shown to be very efficient. 
\end{abstract}
%\noindent{\it Keywords\/}: 
%\submitto{\JSTAT}
\newpage
{\pagestyle{plain}
 \tableofcontents
\cleardoublepage}

\maketitle
\section{Introduction}

Brownian motion is the most popular stochastic process and has a tremendous number of applications in science. In one dimension, a \emph{free} Brownian motion $x(t)$ evolves according to the Langevin equation
\begin{align}
  \dot x(t) =\sqrt{2\,D}\,\eta(t)\,, \label{eq:BMeom}
\end{align} where $D$ is the diffusion coefficient and $\eta(t)$ is an \emph{uncorrelated} Gaussian white noise with zero mean and correlations $\langle \eta(t)\eta(t') \rangle =\delta(t-t')$. In many practical situations, it is necessary to simulate Brownian motion numerically. This can be easily done by discretising the Langevin equation (\ref{eq:BMeom}) over small time increments $\Delta t$:
\begin{align}
  x(t+\Delta t) = x(t) + \sqrt{2\,D}\,\eta(t)\,\Delta t\,,\label{eq:BMeomd}
\end{align}
and drawing at each time step a Gaussian random variable $\sqrt{2\,D}\,\eta(t)\,\Delta t$ with zero mean and variance $2\,D\,\Delta t$. In many applications, such as in the study of foraging animals \cite{Giuggioli05,Randon09,MajumdarCom10,Murphy92,Boyle09}, financial stock markets \cite{Shepp79,Majumdar08}, or in statistical testing \cite{CB2012,Kol1933}, one is only interested in particular trajectories that satisfy some condition. For instance, one can decide to study only \emph{bridge} trajectories which, as their name suggests, are trajectories that start at the origin and return to the origin after a fixed time $t_f$. 
How to generate efficiently such bridge configurations for a Brownian motion? A naive algorithm would be to generate all possible trajectories of Brownian motion up to time $t_f$, starting at the origin, and retain only those that come back to the origin at time $t_f$. Such a naive method is obviously computationally wasteful. This is part of a more general question: how to efficiently sample atypical rare trajectories with a given statistical weight, which is typically very small \cite{BCDG2002,GKP2006,GKLT2011,KGGW2018,Gar2018,Rose21,Rose21area}? In the context of Brownian motion, one can also ask how to generate other constrained Brownian motions, going beyond the bridge. Examples of such constrained Brownian motions include Brownian excursions, Brownian meanders, reflected Brownian motions, etc.  \cite{Yor2000,Majumdar05Ein,MP2010,Dev2010,PY2018}. Fortunately, constrained Brownian motions have been extensively studied and there  exist several methods to sample them \cite{Doob,Pitman,MajumdarEff15,CT2013}. One of them, which is quite powerful and perhaps the easiest relies on writing an effective Langevin equation with an effective force term that implicitly accounts for the constraint \cite{MajumdarEff15,CT2013}. For the Brownian bridge $x_B(t)$, the effective Langevin equation reads \cite{MajumdarEff15,CT2013}
\begin{align}
  \dot x_B(t) = \sqrt{2\,D}\,\eta(t) -\frac{x_B(t)}{t_f-t}\,, \label{eq:BMeomb}
\end{align}
where the subscript $B$ refers to the bridge condition, and the additional term is an effective force term that implicitly accounts for the bridge constraint. The effective Langevin equation (\ref{eq:BMeomb}) can be discretised over time to numerically generate Brownian bridge trajectories with the appropriate statistical weight. The concept of effective Langevin equation is quite robust and can be easily extended to other types of constrained Brownian motions such as excursions, meanders and non-intersecting Brownian motions \cite{CT2013,MajumdarEff15,Orland,Baldassarri2021,Grela2021}. In addition, the concept was recently extended to the case of discrete-time random walks with arbitrary jump distributions, including fat-tailed distributions, and was also shown to be quite a versatile method \cite{DebruyneRW21}.

While for Markov processes, such as the Brownian motion, the effects of constraints (e.g., bridges, excursions, meanders, etc) can be included in an effective Langevin equation (alternatively in effective transition probabilities for discrete-time processes), a similar effective Langevin approach is still lacking for non-Markovian processes which are however abundant in nature \cite{Hanggi1995}. For such processes, there are thus two levels of complexity: (i) the non-Markovian nature of the dynamics indicating temporal correlations in the history of the process and (ii) the effects of the additional geometrical constraints such as the bridge constraint. This two-fold complexity renders the derivation of an effective Langevin equation rather challenging for non-Markovian processes. The goal of this paper is to study an example of a non-Markovian process for which we show that the effective Langevin equation, ensuring the geometric constraints, can be derived exactly.  

Our example of a non-Markovian stochastic process is the celebrated run-and-tumble dynamics of a particle in one dimension, also known as the persistent random walk \cite{kac1974,weiss2002, masoliver2017}, which is of much current interest in the context of active matter \cite{berg08,marchetti13,cates15}. The run-and-tumble particle (RTP) is a simple model that describes self-propelled particles such as the \textit{E. coli} bacteria \cite{berg08}, that are able to move autonomously rendering them inherently different from the standard passive Brownian motion. Active noninteracting particles, including the run-and-tumble model, have been studied extensively in the recent past, both experimentally and theoretically \cite{berg08,marchetti13,cates15,bechinger16,tailleur08}. Even for such  noninteracting systems, a plethora of interesting phenomena have been observed, arising purely from the ``active nature'' of the driving noise. Such phenomena include, e.g., non-trivial density profiles \cite{Bijnens20,Martens12,Basu19,Basu20,Dhar19,Singh20,Santra20,Dean21}, dynamical phase transitions \cite{Doussal20,Gradinego19,Mori21}, anomalous transport properties \cite{Doussal20,Dor19,Demaerel19,Banerjee20}, or interesting first-passage and extremal statistics \cite{Orsingher90,Orsingher95,Lopez14,CinqueF20,CinqueS20,Foong92,Masoliver92,Angelani14,Angelani15,Artuso14,Evans18,Weiss87,Malakar18,Ledoussal19,MoriL20,MoriE20,DebruyneSur21,HartmannConvex20,Singh2019}. 

\begin{figure}[t]
        \centering
        \includegraphics[width=0.4\textwidth]{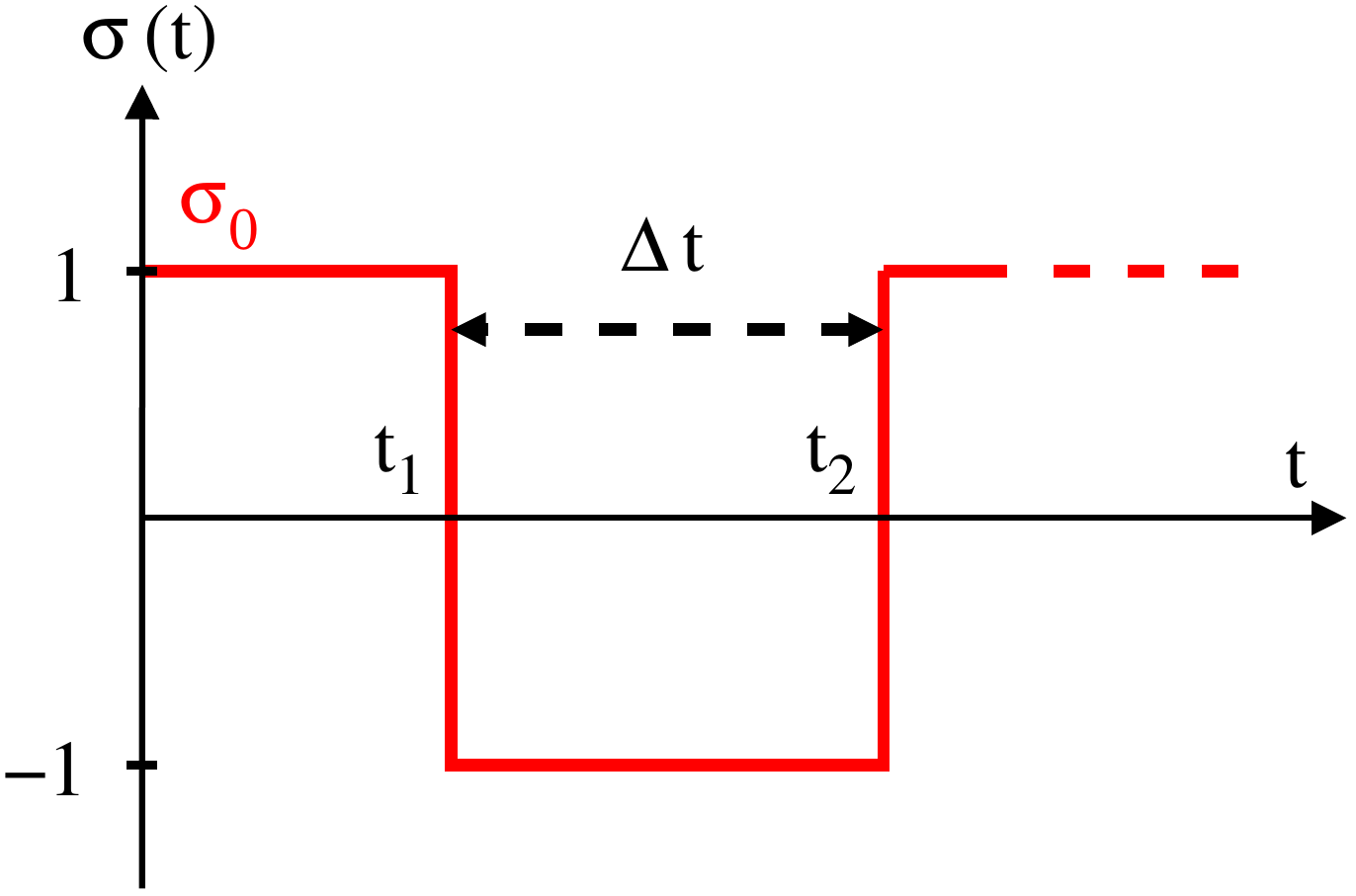}
    \caption{Telegraphic noise $\sigma(t)$ driving the sign of the velocity of the RTP. The signal switches with a constant rate $\gamma$. The time between two consecutive switches $\Delta t$ is drawn independently from an exponential distribution $p(\Delta t)=\gamma \, e^{- \gamma \Delta t}$. The sequence of tumbling times $t_1,\ldots,t_n$ follow a Poisson process of constant rate $\gamma$.}
            \label{fig:telegraphic}
\end{figure}

In its simplest form, a \emph{free} one-dimensional RTP moves (runs) with a fixed velocity $v_0$ in the positive direction during a random time $\Delta t$ drawn from an exponential distribution $p(\Delta t)=\gamma\, e^{-\gamma \Delta t}$ after which it changes direction (tumbles) and goes in the negative direction during another random time. The process continues and the particle performs this run-and-tumble motion indefinitely. The position of the particle $x(t)$ evolves according to the Langevin equation
\begin{align}
  \dot x(t)=v_0\,\sigma(t)\,,\label{eq:eom}
\end{align}
where $\sigma(t)$ is a telegraphic noise that switches between the values $1$ and $-1$ with a \emph{constant} rate $\gamma$ (see figure \ref{fig:telegraphic}). During an infinitesimal time interval $dt$, the particle changes direction with probability $\gamma\, dt$ or remains in the same direction with the complementary probability $1- \gamma\, dt$:
\begin{align}
  \sigma(t+dt) = \left\{\begin{array}{rl}\sigma(t)\,  \quad & \text{with \, prob.~ }=1-\gamma\, dt\, , \\
  -\sigma(t)\,  \quad &\text{with \, prob.~ } =\gamma\, dt\, . \end{array}\right. \label{eq:telegraphic}
\end{align}
Consequently, the time between two consecutive tumbles $\Delta t$ is drawn independently from an exponential distribution $p(\Delta t)=\gamma \, e^{- \gamma \Delta t}$ and the sequence of tumbling times follow a Poisson process with constant rate $\gamma$ (see figure \ref{fig:telegraphic}).
To generate a trajectory $x(t)$ of a free RTP starting from the origin with a given initial velocity
\begin{align}
  x(0)=0\,,\quad \dot x(0) = \sigma_0\,v_0\,,\label{eq:init}
\end{align}
where $\sigma_0=\pm 1$, one simply generates a sequence of tumbling times $t_1,\ldots,t_n$ that follow a \emph{homogeneous} Poisson process of constant rate $\gamma$:
\begin{align}
  t_{m+1} = t_{m} + \Delta t_m\,,\label{eq:tm}
\end{align}
where $\Delta t_m$ are independently drawn from an exponential distribution $p(\Delta t)=\gamma\, e^{-\gamma\,\Delta t}$. Then, the trajectory $x(t)$ of the particle is simply obtained by integrating the equation of motion (\ref{eq:eom}) which yields the piecewise linear function:
\begin{align}
  x(t) = \sigma_0\,v_0\,(-1)^n\, (t-t_n)+\sum_{m=0}^{n-1} \sigma_0\, v_0\, (-1)^m \, (t_{m+1}-t_{m}) \,,\label{eq:xt}
\end{align}
where $n$ is such that $t_n$ is the latest tumbling time before $t$, i.e. such that $t_n<t<t_{n+1}$. The sum in (\ref{eq:xt}) accounts for all complete runs that happened before $t$ and the first term corresponds to the last run that is not yet completed at time $t$. This sampling method works well to generate \emph{free} run-and-tumble trajectories. However, as in the case of Brownian motion, some applications require to only sample specific trajectories, such as bridge trajectories where, in addition to satisfy the initial condition (\ref{eq:init}), the particle must also return to the origin after a fixed time $t_f$ with a given velocity $\sigma_f\,v_0$:
\begin{align}
  x(t_f)=0\,,\quad \dot x(t_f) = \sigma_f\, v_0\,,\label{eq:final}
\end{align}
where $\sigma_f =\pm 1$. Note that the final position need not necessarily be the origin but any fixed point in space -- here for simplicity we only consider the case where the final position coincides with the origin. One possible application of run-and-tumble bridge trajectories is in the context of animal foraging, where animals typically return to their nest after a fixed time, and one could study the persistence and memory effects in their trajectories
 \cite{Giuggioli05,Randon09,MajumdarCom10,Murphy92,Boyle09}. 
Unfortunately, as in the case of Brownian motion, obtaining realisations of bridge trajectories using the free sampling method would be computationally wasteful. As argued in the introduction, 
one needs an efficient algorithm to generate run-and-tumble bridge trajectories, in a similar spirit as the effective Langevin equation (\ref{eq:BMeomb}) for Brownian motion. 
In this paper, we derive an exact effective Langevin equation for RTPs to generate bridge trajectories efficiently. We show that the effective process, that automatically takes care of the bridge constraints (\ref{eq:init}) and (\ref{eq:final}) can be written as
\begin{align}
  \dot x(t)=v_0\,\sigma^*(x,\dot x,t\,|\,\sigma_0,t_f,\sigma_f)\,,\label{eq:effeom}
\end{align}
where $\sigma^*(x,\dot x,t\,|\,\sigma_0,t_f,\sigma_f)$ is now an effective telegraphic noise that switches between the values $1$ and $-1$ with a space-time dependent rate $ \gamma^*(x,\dot x,t\,|\,\sigma_0,t_f,\sigma_f)$, which we compute exactly (\ref{eq:effB}). Finally, we show how to extend the method to other types of constrained RTP trajectories, such as the excursion (a bridge RTP that is additionally constrained to remain above the origin) and the meander (where the RTP is constrained not to cross the origin and with a free end point). We illustrate our method by numerical simulations (the code is available as a Python notebook in \cite{github}).

The rest of the paper is organised as follows. In section \ref{sec:bridge}, we present the derivation of the effective Langevin equation for the bridge RTP and derive the effective tumbling rate that accounts for the bridge constraint. In section \ref{sec:gen}, we generalise the effective Langevin equation to the case of other constrained run-and-tumble trajectories such as the excursion and the meander and derive their effective tumbling rates. Finally, in section \ref{sec:sum}, we conclude and provide perspectives for further research. Some useful results on the run-and-tumble process are recalled in \ref{app:prop}.

\section{Generating run-and-tumble bridges}
\label{sec:bridge}
\begin{figure}[t]
        \centering
        \includegraphics[width=0.4\textwidth]{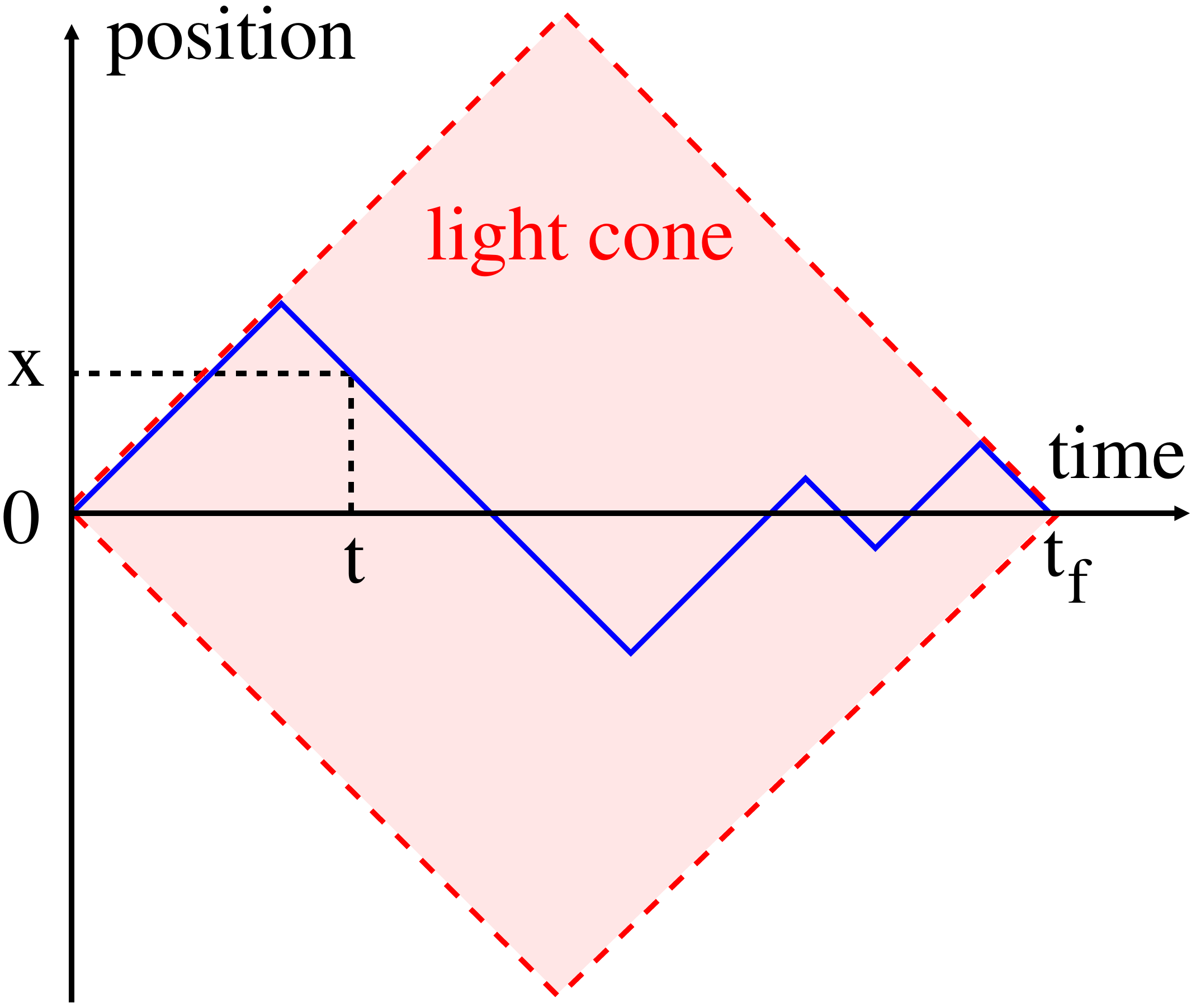}
    \caption{A sketch of a run-and-tumble bridge trajectory that starts at the origin with a positive velocity $\dot x=+v_0$ and returns to the origin at a fixed time $t_f$ with a negative velocity $\dot x=-v_0$.  Due to the Markov property in the extended phase space $(x,\dot x)$, the bridge trajectory can be decomposed into two independent parts:  a left part over the time interval $[0,t]$, where the particle freely moves from the point $(0,+v_0)$ to the point $(x,-v_0)$ at time $t$ and a right part over the time interval $[t,t_f]$, where it moves from the point $(x,-v_0)$ at time $t$ to the point $(0,-v_0)$ at time $t_f$. The combination of the finite velocity of the particle and the bridge condition induces a double sided light cone in which the particle must remain (shaded red region).}
            \label{fig:bridges}
\end{figure}

The derivation of the effective Langevin equation for the bridge RTP follows similar ideas to the ones developed for continuous and discrete time Markov processes \cite{MajumdarEff15,DebruyneRW21}. The key point is that the free run-and-tumble process, though non-Markovian in the $x$-coordinate, becomes Markovian 
in the phase space $(x,\dot x)$. Therefore, a bridge trajectory satisfying the initial and final conditions (\ref{eq:init})-(\ref{eq:final}) can be decomposed into two independent paths over the time intervals $[0,t]$ and $[t,t_f]$ (see figure \ref{fig:bridges}).
 As a result, the bridge probability distribution $P_B(x,t,\sigma\,|\,\sigma_0,t_f,\sigma_f)$ to find the particle at $x$ with a velocity $\dot x =\sigma\,v_0$ at time $t$ given that it satisfies the bridge conditions (\ref{eq:init})-(\ref{eq:final}) can be decomposed as a simple product
\begin{align}
  P_B(x,t,\sigma\,|\,\sigma_0,t_f,\sigma_f)=\frac{P(x,t,\sigma\,|\,\sigma_0)\,Q(x,t_f-t,\sigma\,|\,\sigma_f)}{P(x=0,t_f,\sigma_f\,|\,\sigma_0)}\,,\label{eq:Pb}
\end{align}
where the subscript $B$ refers to ``bridge''. The first term $P(x,t,\sigma|\sigma_0)$ in (\ref{eq:Pb}) accounts for the first path over $[0,t]$ and is the probability density of the free particle to be located at position $x$ at time $t$ with velocity $\dot x=\sigma\, v_0$ given that it started at the origin with velocity $\dot x=\sigma_0\, v_0$. This is usually referred to as the forward propagator. The second term $Q(x,t,\sigma\,|\,\sigma_f)$ is the probability density of the free particle to reach the origin at time $t$ with velocity $\dot x=\sigma_f\,v_0$ given that it started at $x$ with velocity $\dot x=\sigma\, v_0$. We will refer to it as the backward propagator. The denominator in (\ref{eq:Pb}) is a normalisation factor that accounts for all the bridge trajectories such that $\int_{-\infty}^{\infty} dx \sum_{\sigma=\pm}P_B(x,t,\sigma\,|\,\sigma_0,t_f,\sigma_f)=1$.

Using Markov properties, one can see that the free forward propagator $P(x,t,\sigma|\sigma_0)$ and backward propagator $Q(x,t,\sigma\,|\,\sigma_f)$ evolve according to Fokker-Plank equations. For conciseness, we will drop the conditional dependence in the differential equations below and use the shorthand notation $P(x,t,\sigma)\equiv P(x,t,\sigma|\sigma_0)$, $Q(x,t,\sigma)\equiv Q(x,t,\sigma\,|\,\sigma_f)$. To obtain the Fokker-Plank equations for the forward propagator, let us consider an infinitesimal time interval $[t-dt,t]$ and suppose that the particle is located at $x$ at time $t$ with velocity $\dot x=\sigma\,v_0$. In the time interval $[t-dt,t]$, we see from the telegraphic equation (\ref{eq:telegraphic}) that the particle either traveled with velocity $\dot x=\sigma \,v_0$ from $x-\sigma \,v_0\,dt$ to $x$ or tumbled with velocity $\dot x=-\sigma \,v_0$ and remained at $x$. The first event happens with probability $1-\gamma\,dt$ and the second event happens with the complementary probability $\gamma\,dt$. We can now write the following equation for the forward propagator
\begin{align}
  P(x,t,\sigma) = (1-\gamma\,dt)\,P(x-\sigma \,v_0\,dt,t-dt,\sigma) + \gamma\,dt \,\,P(x,t-dt,-\sigma)\,.\label{eq:Pdt}
\end{align}
Expanding (\ref{eq:Pdt}) to first order in $dt$ and writing separate equations for $\sigma=+1$ and $\sigma=-1$, we find that $P(x,t,\sigma)$ satisfies a set of two coupled equations, called the \emph{forward} Fokker-Plank equations:
\begin{subequations}
\begin{align}
   \partial_t P(x,t,+)&=- v_0\,\partial_x P(x,t,+)-\gamma\, P(x,t,+)+\gamma\, P(x,t,-)\,,\\
    \partial_t P(x,t,-)&= +v_0\,\partial_x P(x,t,-)-\gamma\, P(x,t,-)+\gamma\, P(x,t,+)\,.
\end{align}
\label{eq:P}
\end{subequations}
The forward propagator $P(x,t,\sigma|\sigma_0)$ of the free particle can be obtained analytically by solving the differential equations (\ref{eq:P}) on the real line along with the initial condition $P(x,t=0,\sigma|\sigma_0)=\delta_{\sigma,\sigma_0}\,\delta(x)$. To obtain the Fokker-Plank equations for the backward propagator, we instead consider an infinitesimal time interval $[0,dt]$ and suppose that the particle is initially located at $x$ at time $t=0$ with velocity $\dot x=\sigma\,v_0$. In the time interval $[0,dt]$, the particle either traveled with velocity $\dot x=\sigma \,v_0$ to $x+\sigma \,v_0\,dt$ or tumbled to a velocity $\dot x=-\sigma \,v_0$ and remained at $x$. After either of these two events, the particle must reach the origin in a time $t-dt$. Therefore, we can write the following equation for the backward propagator
\begin{align}
  Q(x,t,\sigma) = (1-\gamma\,dt)\,Q(x+\sigma \,v_0\,dt,t-dt,\sigma) + \gamma\,dt \,\,Q(x,t-dt,-\sigma)\,,\label{eq:Qdt}
\end{align}
which, after expanding to first order in $dt$, gives the \emph{backward} Fokker-Plank equations:
\begin{subequations}
\begin{align}
 - \partial_t Q(x,t,+)= +v_0\,\partial_x Q(x,t,+)-\gamma \,Q(x,t,+)+\gamma\, Q(x,t,-)\,,\\
 - \partial_t Q(x,t,-)=- v_0\,\partial_x Q(x,t,-)-\gamma\, Q(x,t,-)+\gamma\, Q(x,t,+)\,.
\end{align}
\label{eq:Q}
\end{subequations}
The backward propagator $Q(x,t,\sigma|\sigma_f)$ of the free particle can be obtained analytically by solving the differential equations (\ref{eq:Q}) on the real line along with the initial condition $Q(x,t=0,\sigma|\sigma_f)=\delta_{\sigma,\sigma_f}\,\delta(x)$. The derivation can be found in e.g. \cite{DebruyneSur21} and the results are recalled in \ref{app:prop}. It is now easy to show that the bridge propagator $ P_B(x,t,\sigma\,|\,\sigma_0,t_f,\sigma_f)$ defined in (\ref{eq:Pb}) in terms of $P$ and $Q$ satisfies a similar set of Fokker-Plank equations. Omitting the conditional dependence for conciseness, we find that the bridge propagator satisfies the effective Fokker-Plank equations
\begin{subequations}
\begin{align}
   \partial_t P_B(x,t,+) &= - v_0\partial_xP_B(x,t,+)-    \gamma_B^*(x,+,t)  P_B(x,t,+) +    \gamma_B^*(x,-,t) P_B(x,t,-)\,,\\[1em]
    \partial_t P_B(x,t,-) &= + v_0\partial_xP_B(x,t,-)- \gamma_B^*(x,-,t) P_B(x,t,-)+ \gamma_B^*(x,+,t) P_B(x,t,+)\,,
\end{align}
\label{eq:effFPb}
\end{subequations}
where the transition rates are now space-time dependent:
\begin{subequations}
\begin{align}
  \gamma_B^*(x,\dot x=+v_0,t\,|\,\sigma_0,t_f,\sigma_f) &= \gamma\, \frac{Q(x,\tau,-\,|\,\sigma_f)}{Q(x,\tau,+\,|\,\sigma_f)}\,,\\[1em]
  \gamma_B^*(x,\dot x=-v_0,t\,|\,\sigma_0,t_f,\sigma_f) &= \gamma\, \frac{Q(x,\tau,+\,|\,\sigma_f)}{Q(x,\tau,-\,|\,\sigma_f)}\,,
\end{align}
\label{eq:effBQ}
\end{subequations}
where $\tau=t_f-t$ and $Q$ is the free backward propagator satisfying the backward Fokker-Plank equations (\ref{eq:Q}). One can easily check that the effective equations (\ref{eq:effFPb}) conserve the probability current such that the bridge propagator is indeed normalised to unity $\int_{-\infty}^{\infty} dx \sum_{\sigma=\pm}P_B(x,t,\sigma\,|\,\sigma_0,t_f,\sigma_f)=1$. Physically, the effective tumbling rate is the free tumbling rate that is modified in such a way that tumbling events that bring the particle closer to the origin are more likely to happen. Using the expression of the free backward propagator (recalled in \ref{app:prop}), we find the exact expressions of the transition rates. For example, when $\sigma_0=+1$ and $\sigma_f=-1$, we get
\begin{subequations}
\begin{align}
  \gamma_B^*(x,\dot x=+v_0,t\,|\,+,t_f,-) &= 2\,\gamma\,\delta[f(\tau,x)]+\,\gamma\,\sqrt{\frac{g(\tau,x)}{f(\tau,x)}} \frac{I_1[ h(\tau,x)]}{I_0[ h(\tau,x)]}\,,\\[1em]
  \gamma_B^*(x,\dot x=-v_0,t\,|\,+,t_f,-) &= \gamma\, \frac{1}{2\,\delta[f(\tau,x)]+\sqrt{\frac{g(\tau,x)}{f(\tau,x)}} \frac{I_1[ h(\tau,x)]}{I_0[ h(\tau,x)]}}\,,
\end{align}
\label{eq:effB}
\end{subequations}
where $\tau=t_f-t$. In the expressions (\ref{eq:effB}), $I_0(z)$ and $I_1(z)$ denote the modified Bessel functions while the functions $f$, $g$, and $h$ are defined as   
\begin{align}
    f(t,x) = \gamma\,t-\frac{\gamma\,x}{v_0}\,,\quad g(t,x)= \gamma\,t+\frac{\gamma\,x}{v_0}\,, \quad h(t,x)=\sqrt{f(t,x)\,g(t,x)}\,. \label{eq:fgh}
  \end{align}  
  The Dirac delta terms in the effective rates (\ref{eq:effB}) enforce the particle to remain in the double sided light cone defined as (see figure \ref{fig:bridges})
 \begin{align}
   \left\{\begin{array}{ll} |x|\leq v_0\, t\,, &\text{when }\, 0\leq t \leq \frac{t_f}{2}\, ,\\ 
   |x|\leq v_0 \,(t_f-t) \,, &\text{when }\, \frac{t_f}{2}\leq t \leq t_f\, ,\label{eq:lc}
\end{array} \right.
 \end{align}
which is a natural boundary induced by the combination of the finite velocity of the particle along with the bridge constraint. In practice, when performing numerical simulations, these Dirac delta terms can be safely removed from the effective tumbling rates and can be replaced by hard constraints such that the particle must remain in the double sided light cone (\ref{eq:lc}). 

\begin{figure}[t]
\subfloat{%
 \includegraphics[width=0.5\textwidth]{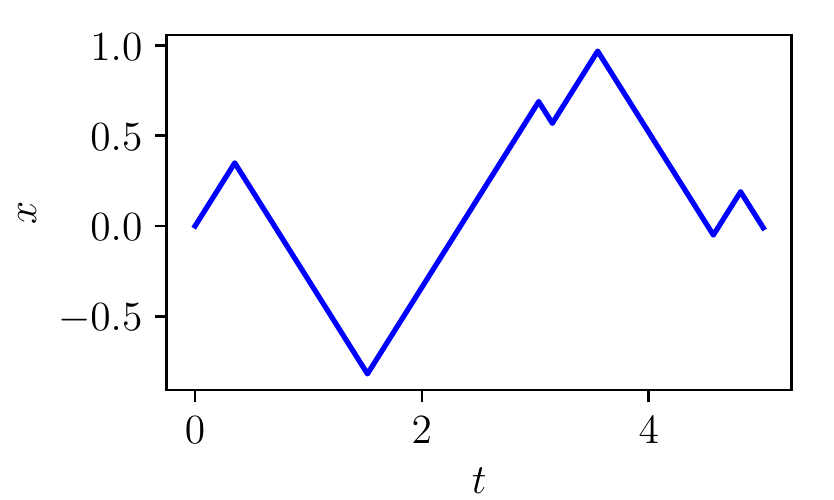}%
}\hfill
\subfloat{%
  \includegraphics[width=0.5\textwidth]{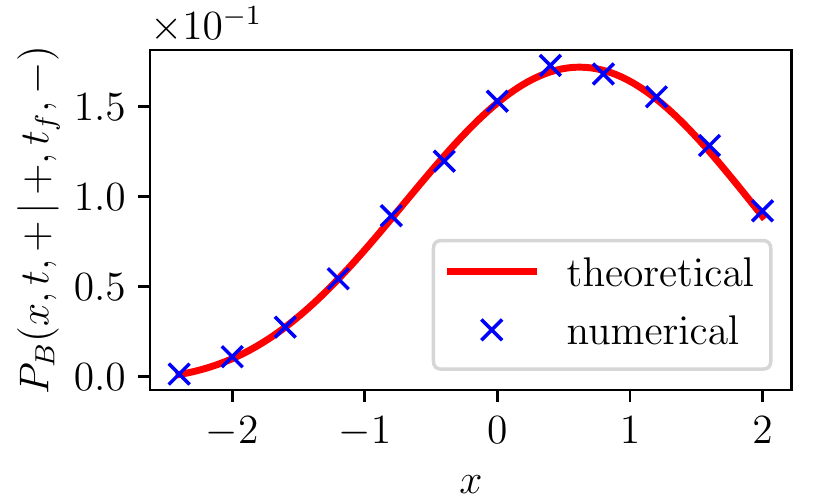}%
}\hfill
\caption{\textbf{Left panel:} A typical bridge trajectory of a RTP starting at the origin with a positive velocity $\dot x=+v_0$ and returning to the origin after a time $t_f=5$ with a negative velocity $\dot x=-v_0$. The trajectory was generated using the effective tumbling rates (\ref{eq:effB}). \textbf{Right panel:} Position distribution at $t=t_f/2$ for a RTP starting at the origin with a positive velocity $\dot x=+v_0$ and returning to the origin after a time $t_f=5$ with a negative velocity $\dot x=-v_0$. The position distribution $P_B(x,t,+\,|\,+,t_f,-)$ obtained numerically by sampling from the effective tumbling rates (\ref{eq:effB}) is compared with the theoretical prediction (\ref{eq:Pbridge}). The agreement is excellent. Note that the Dirac delta function in (\ref{eq:Pbridge}) is not shown to fit the data within the limited window size.}\label{fig:bridge}
\end{figure}

By comparing the effective Fokker-Plank equations for the bridge propagator (\ref{eq:effFPb}) with the ones for the free propagator (\ref{eq:P}), one can see that the bridge constraint is encoded in the space-time dependency of the tumbling rates and lead to the effective Langevin equation (\ref{eq:effeom}) with a space-time dependent telegraphic noise presented in the introduction. RTPs with space and time dependent tumbling rates are relatively easy to simulate and there have been quite a few recent studies on them~\cite{Doussal20,Dor19,Singh20,Angelani14}. 
Unlike these models where the space and time dependency of the tumbling rates are ``put in by hand'', here we see from first principle how geometric constraints, such as the bridge condition, naturally generates space-time dependent tumbling rates. To generate trajectories of RTPs with space-time dependent tumbling rates, one proceeds as follows.  
Instead of generating a sequence of tumbling times that follow a \emph{homogeneous} Poisson process with constant rate $\gamma$, as presented in the introduction, one needs to generate a sequence of times that follow a \emph{non homogeneous} Poisson process with a variable rate. There exist several methods to generate non homogeneous Poisson processes (see \cite{Lewis79} for a review). A quick and simple method is to discretise the effective equation (\ref{eq:effeom}) over small time increments $\Delta t$ which, omitting the conditional dependence, writes
\begin{align}
  x_B(t+\Delta t) = x_B(t) + v_0\,\Delta t\,\sigma_B^*(x_B,\dot x_B,v_0,t)\,,\label{eq:eomd}
\end{align}
and to evolve the telegraphic signal according to
\begin{align}
 \sigma_B^*(x_B,\dot x_B,v_0,t+\Delta t) = \left\{\begin{array}{rl}\sigma_B^*(x_B,\dot x_B,v_0,t)\, \quad  & \text{with \, prob.~ }=1-\gamma_B^*(x_B,\dot x_B,t)\, \Delta t\, , \\
  -\sigma_B^*(x_B,\dot x_B,v_0,t)\, \quad &\text{with \, prob.~ } =\gamma_B^*(x_B,\dot x_B,t)\, \Delta t\, . \end{array}\right. \label{eq:telegraphicN}
\end{align}
This method is very simple to implement but nevertheless requires to choose the time increments $\Delta t$ sufficiently small such that the switching probabilities in (\ref{eq:telegraphicN}) do not exceed unity, which can be an issue if one is interested in regimes close to the light cone structure where the effective rates become large and might require more advanced sampling techniques \cite{Lewis79}. Nevertheless, this method effectively generates run-and-tumble bridge trajectories and works well in practice (see left panel in figure \ref{fig:bridge}).  In the right panel in figure \ref{fig:bridge}, we computed numerically the probability distribution of the position at some intermediate time $t=t_f/2$, by generating bridge trajectories from the effective tumbling rates (\ref{eq:effB}) and compared it to the theoretical position distribution for the bridge propagator which can be easily computed by substituting the free forward and backward propagators (recalled in the \ref{app:prop}) in the expression of the bridge propagator in (\ref{eq:Pb}):
   \begin{subequations}
\begin{align}
P_B(x,t,-\,|\,+,t_f,-)&=\frac{\gamma}{2\,v_0}\frac{I_0[h(t,x)]}{I_0[\gamma\, t_f]}\,\left(2\,\delta[f(\tau,x)]+\sqrt{\frac{g(\tau,x)}{f(\tau,x)}}I_1[h(\tau,x)]\right)\,, \\
P_B(x,t,+\,|\,+,t_f,-)&=P_B(x,\tau,-\,|\,+,t_f,-)\,,
\end{align}
\label{eq:Pbridge}
\end{subequations}
where $\tau=t_f-t$. In the expressions (\ref{eq:Pbridge}), $I_0(z)$ and $I_1(z)$ denote the modified Bessel functions. As can be seen in figure \ref{fig:bridge}, the agreement is excellent.
Note that in the diffusive limit when
  \begin{align}
    v_0\rightarrow\infty\,,\quad \gamma\rightarrow\infty \,,\quad \text{with } D\equiv\frac{v_0^2}{2\gamma} \text{ fixed}\,,\label{eq:bml}
  \end{align}
  where $D$ is the effective diffusion coefficient, the effective tumbling rates (\ref{eq:effB}) both become the same constant $\gamma$ which is independent of $x$ and $t$. The signature of the bridge constraint can be found in the second order term of this limit which gives 
 \begin{subequations}
\begin{align}
 \gamma_B^*(x,\dot x=+v_0,t\,|\,+,t_f,-)  &\sim \gamma+\frac{x}{\tau\,\sqrt{2\,D}}\,\gamma^{\frac{1}{2}}+O(\gamma^{-1}),\\
 \gamma_B^*(x,\dot x=-v_0,t\,|\,+,t_f,-)  &\sim \gamma-\frac{x}{\tau\,\sqrt{2\,D}}\,\gamma^{\frac{1}{2}}+O(\gamma^{-1})\,.
\end{align} 
\label{eq:effgbm}
 \end{subequations}
Note that one needs to retain the subleading terms up to order $O(\sqrt{\gamma})$ in order to capture the nontrivial $x$-dependence, which indeed ensures the bridge condition. Upon inserting these rates in the effective Fokker-Plank equations (\ref{eq:effFPb}) and solving for $P_B(x,t)\equiv P_B(x,t,+)+P_B(x,t,-)$ by adding and subtracting the two equations, we find that the first order terms in the tumbling rates cancel out and we recover the well-known effective Fokker-Plank equation for Brownian motion
  \begin{align}
  \partial_t P_B(x,t) = D\partial_x[\partial_x P_B(x,t) - 2 P_B(x,t)\partial_x\ln(Q(x,\tau))]\,,\label{eq:DiffB}
  \end{align}
 where $\tau=t_f-t$ and $Q(x,\tau)=\frac{1}{\sqrt{4\pi D\tau}}e^{-x^2/4D\tau}$ is the free Brownian backward propagator. This Fokker-Plank equation leads to the effective Langevin equation (\ref{eq:BMeomb}) that generates Brownian bridges presented in the introduction.

%
% \paragraph{Maximum of the bridge} The joint distribution $p(x_\text{max},t_\text{max})$ of the position $x_\text{max}$ and the time $t_\text{max}$ of the maximum can be obtained by noting that it is composed of two first-passage paths (see figure ):
% \begin{align}
%   p(x_\text{max},t_\text{max}) = \frac{1}{v_0}\,F(t_\text{max}|x_\text{max},-\sigma_0)\,F(t_f-t_\text{max}|x_\text{max},\sigma_f)\,
% \end{align}

\section{Generalisation to other constrained run-and-tumble trajectories} 
\label{sec:gen}
In the previous section, we obtained effective tumbling rates to generate bridge run-and-tumble trajectories. In this section, we generalise the method to other types of constrained run-and-tumble trajectories, namely excursions and meanders.
\subsection{Generating run-and-tumble excursions} 
\begin{figure}[t]
        \centering
        \includegraphics[width=0.4\textwidth]{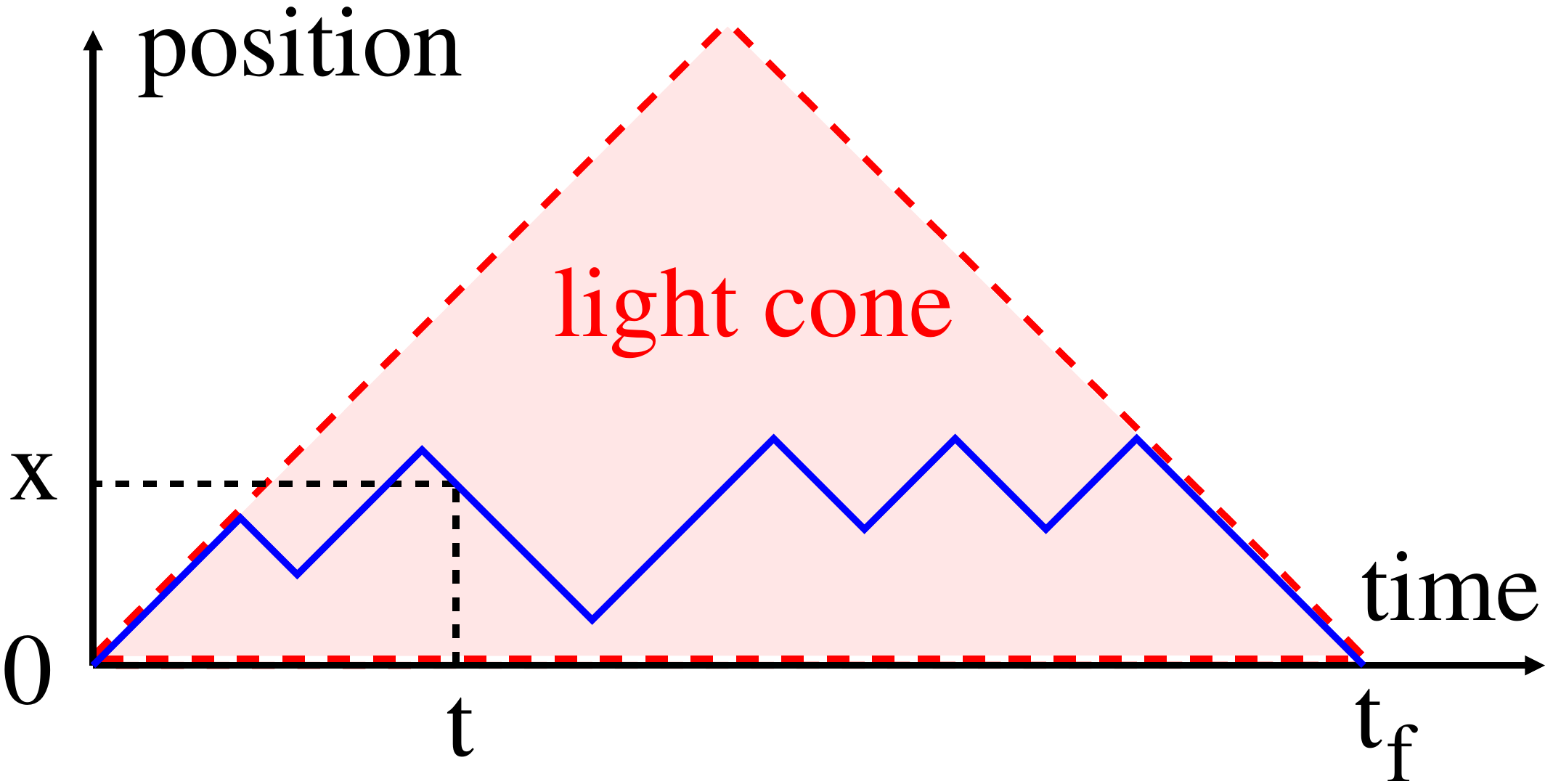}
    \caption{A sketch of a run-and-tumble excursion trajectory that starts at the origin and returns to the origin at a fixed time $t_f$ while remaining positive.  Due to the Markov property in the extended phase space $(x,\dot x)$, the excursion trajectory can be decomposed into two independent parts:  a left part over the interval $[0,t]$, where the particle moves from the point $(0,+v_0)$ to the point $(x,-v_0)$ at time $t$ while staying positive and a right part over the interval $[t,t_f]$, where it moves from the point $(x,-v_0)$ at time $t$ to the point $(0,-v_0)$ at time $t_f$ while staying positive. The combination of the finite velocity of the particle and the excursion condition induces a positive double sided light cone in which the particle must remain (shaded red region).}
            \label{fig:excursions}
\end{figure}

An excursion is a bridge trajectory that is further constrained to remain above the origin. The particle must start from the origin $x_0=0$, necessarily in the state $\sigma_0=+1$, and return to the origin at the time $t_f$, necessarily in the state $\sigma_f=-1$, while never crossing the origin:
  \begin{align}
x(0)=x(t_f)=0\,,\quad \dot x(0)=+v_0\,,\quad x(t')\geq 0\quad \forall t' \in [0,\,t_f]\,,\quad \dot x(t_f)=-v_0\,.\label{eq:exc}
  \end{align}
Similarly to the bridge propagator (\ref{eq:Pb}), the propagator for an excursion can be written as (see figure \ref{fig:excursions})
\begin{align}
   P_E(x,t,\sigma\,|\,t_f)=\frac{P_{\text{absorbing}}(x,t,\sigma)\,Q_{\text{absorbing}}(x,t_f-t,\sigma)}{P_{\text{absorbing}}(x=0,t_f)}\,,\label{eq:Pe}
\end{align} where the subscript $E$ refers to ``excursion'', and $P_{\text{absorbing}}$ and $Q_{\text{absorbing}}$ are now the forward and backward propagator of the free RTP in the presence of an absorbing boundary located at the origin.
 They satisfy the set of Fokker-Plank equations (\ref{eq:P}) and (\ref{eq:Q}) that must now be solved on the half line with the initial condition $P_{\text{absorbing}}(x,t\!=\!0,\sigma)=\delta_{\sigma,+}\delta(x)$ and $Q_{\text{absorbing}}(x,t\!=\!0,\sigma)=\delta_{\sigma,-}\delta(x)$. The boundary conditions at $x=0$ can be obtained by looking at the differential forms (\ref{eq:Pdt})-(\ref{eq:Qdt}) and are found to be $P_{\text{absorbing}}(x\!=\!0,t,+)=0$ and $Q_{\text{absorbing}}(x\!=\!0,t,-)=0$. Following the steps in the previous section, we find that the analog of the effective tumbling rates (\ref{eq:effBQ}) are given by
\begin{subequations}
\begin{align}
   \gamma_E^*(x,\dot x=+v_0,t\,|\,t_f) &= \gamma\, \frac{Q_{\text{absorbing}}(x,\tau,-)}{Q_{\text{absorbing}}(x,\tau,+)}\,,\\[1em]
  \gamma_E^*(x,\dot x=-v_0,t\,|\,t_f) &= \gamma\, \frac{Q_{\text{absorbing}}(x,\tau,+)}{Q_{\text{absorbing}}(x,\tau,-)}\,,
\end{align}
  \label{eq:gEQ}
\end{subequations}
where $\tau=t_f-t$ and $Q_{\text{absorbing}}$ is the backward propagator of the free particle in the presence of an absorbing boundary. Using its expression (recalled in \ref{app:prop}), we find the exact expressions of the transition rates:
  \begin{subequations}
  \begin{align}
    \gamma_E^*(x,\dot x=+v_0,t\,|\,t_f) &=  2\,\frac{\gamma v_0\tau}{ x}\,\delta[f(\tau,x)]+\frac{\gamma^2\, x}{v_0}\,\sqrt{\frac{g(\tau,x)}{f(\tau,x)}}\frac{I_1[h(\tau,x)]}{\frac{\gamma x}{v_0}I_0[h(\tau,x)]+\sqrt{\frac{f(\tau,x)}{g(\tau,x)}}I_1[h(\tau,x)]}\,,\\
      \gamma_E^*(x,\dot x=-v_0,t\,|\,t_f) &= \gamma\, \frac{1}{2\,\frac{v_0\tau}{ x}\,\delta[f(\tau,x)]+\frac{\gamma\, x}{v_0}\,\sqrt{\frac{g(\tau,x)}{f(\tau,x)}}\frac{I_1[h(\tau,x)]}{\frac{\gamma x}{v_0}I_0[h(\tau,x)]+\sqrt{\frac{f(\tau,x)}{g(\tau,x)}}I_1[h(\tau,x)]}}\,, 
  \end{align}
  \label{eq:effE}
  \end{subequations}
where $\tau=t_f-t$. In the expressions (\ref{eq:effE}), $I_0(z)$ and $I_1(z)$ denote the modified Bessel functions while the functions $f$, $g$, and $h$ are defined in (\ref{eq:fgh}). As in the bridge case, the Dirac delta terms in the effective rates (\ref{eq:effE}) enforce the particle to remain in the positive double sided light cone defined as (see figure \ref{fig:excursions})
 \begin{align}
   \left\{\begin{array}{ll} 0\leq x\leq v_0\, t\,, &\text{when }\, 0\leq t \leq \frac{t_f}{2}\, ,\\ 
   0\leq x\leq v_0 \,(t_f-t) \,, &\text{when }\, \frac{t_f}{2}\leq t \leq t_f\, ,\label{eq:plc}
\end{array} \right.
 \end{align}
which is a natural boundary induced by the combination of the finite velocity of the particle along with the excursion constraint. In practice, when performing numerical simulations, these Dirac delta terms can be safely removed from the effective tumbling rates and be replaced by hard constraints such that the particle must remain in the positive double sided light cone (\ref{eq:plc}). 

\begin{figure}[t]
\subfloat{%
 \includegraphics[width=0.5\textwidth]{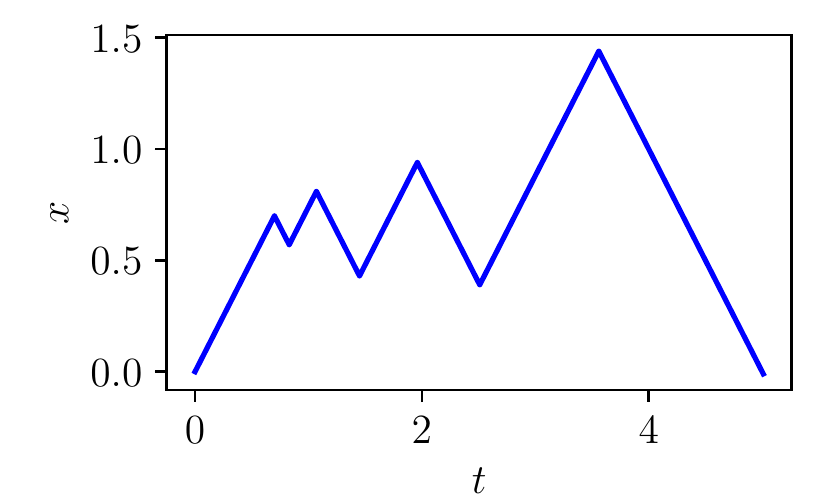}%
}\hfill
\subfloat{%
  \includegraphics[width=0.5\textwidth]{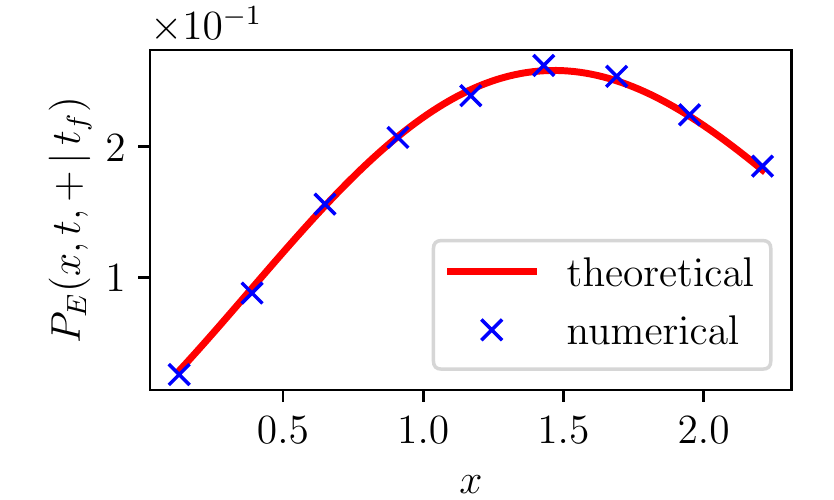}%
}\hfill
\caption{\textbf{Left panel:} A typical excursion trajectory of a RTP starting at the origin and returning to the origin after a time $t_f=5$ while remaining positive. The trajectory was generated using the effective tumbling rates (\ref{eq:effE}). \textbf{Right panel:} Position distribution at $t=t_f/2$ for a RTP starting at the origin and returning to the origin after a time $t_f=5$ while remaining positive. The position distribution $P_E(x,t,+\,|\,t_f)$ obtained numerically by sampling from the effective tumbling rates (\ref{eq:effE}) is compared with the theoretical prediction (\ref{eq:Pexcursion}). The agreement is excellent. Note that the Dirac delta function in (\ref{eq:Pexcursion}) is not shown to fit the data within the limited window size.}\label{fig:excursion}
\end{figure}

The effective rates (\ref{eq:effE}) generate run-and-tumble excursion trajectories (see left panel in figure \ref{fig:excursion}).  In the right panel in figure \ref{fig:excursion}, we computed numerically the probability distribution of the position at some intermediate time $t=t_f/2$, by generating excursion trajectories from the effective tumbling rates (\ref{eq:effE}) and compared it to the theoretical position distribution for the excursion propagator which can be easily computed by substituting the forward and backward propagators of a free particle in the presence of an absorbing boundary (recalled in the \ref{app:prop}) in the expression of the excursion propagator in (\ref{eq:Pe}):
   \begin{subequations}
\begin{align}
P_E(x,t,+) &= \frac{\gamma\,t_f}{I_1[\gamma t_f]\,(v_0\,\tau+x)}\left(\frac{\gamma x}{v_0}\,I_0[h(\tau,x)]+\sqrt{\frac{f(\tau,x)}{g(\tau,x)}}I_1[h(\tau,x)]\right)\nonumber\\
 &\quad \times \left(\delta[f(t,x)]+\frac{\gamma\,x}{v_0}\frac{1}{\sqrt{f(t,x)\,g(t,x)}} I_1[h(t,x)]\right)\,,\\
P_E(x,t,-)&=P_E(x,\tau,+)\,,
\end{align}
\label{eq:Pexcursion}
\end{subequations}
where $\tau=t_f-t$. In the expressions (\ref{eq:Pexcursion}), $I_0(z)$ and $I_1(z)$ denote the modified Bessel functions while the functions $f$, $g$, and $h$ are defined in (\ref{eq:fgh}). As can be seen in figure \ref{fig:excursion}, the agreement is excellent.
As in the bridge case, we can compute the diffusive limit (\ref{eq:bml}) of the effective rates (\ref{eq:effE}) to find that they take a rather simple form
\begin{subequations}
\begin{align}
   \gamma_E^*(x,\dot x=+v_0,t\,|\,t_f) &\sim \gamma + \left(\frac{x}{\tau\,\sqrt{2D}}-\frac{\sqrt{2D}}{x}\right)\,\gamma^{\frac{1}{2}}+O(\gamma^{-1})\,,\\
   \gamma_E^*(x,\dot x=-v_0,t\,|\,t_f) &\sim\gamma - \left(\frac{x}{\tau\,\sqrt{2D}}-\frac{\sqrt{2D}}{x}\right)\,\gamma^{\frac{1}{2}}+O(\gamma^{-1})\,,
\end{align} 
\label{eq:gEbm}
\end{subequations}
which, upon inserting in the effective Fokker-Plank equations (\ref{eq:effFPb}) gives back the effective Langevin equation for Brownian excursions \cite{MajumdarEff15}.

\subsection{Generating run-and-tumble meanders} 
\begin{figure}[t]
        \centering
        \includegraphics[width=0.4\textwidth]{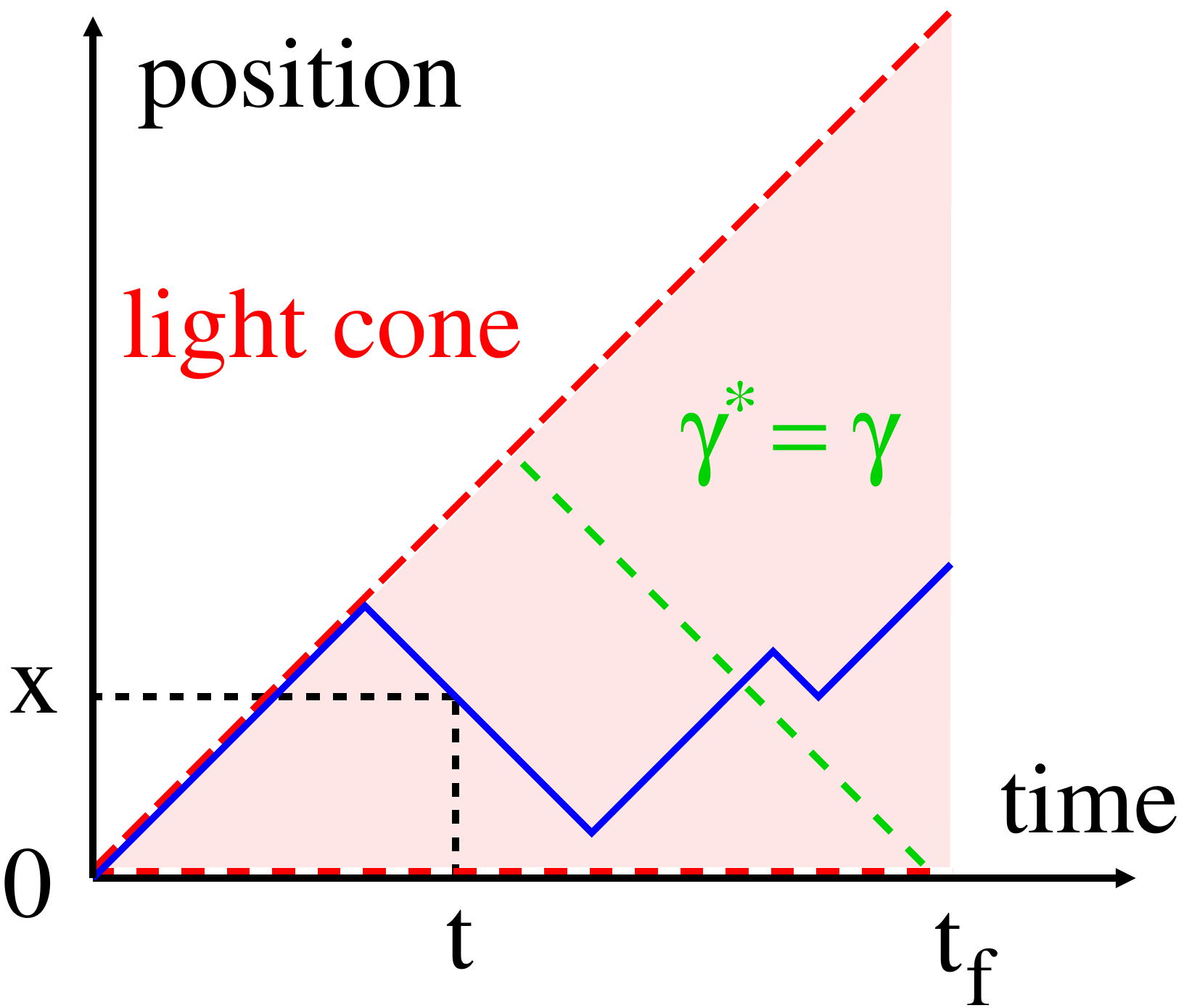}
    \caption{A sketch of a run-and-tumble meander trajectory that starts at the origin and remains positive up to a time $t_f$.  Due to the Markov property in the extended phase space $(x,\dot x)$, the meander trajectory can be decomposed into two independent parts:  a left part over the interval $[0,t]$, where the particle moves from the point $(0,+v_0)$ to the point $(x,-v_0)$ at time $t$ while staying positive, and a right part over the interval $[t,t_f]$, where it moves from the point $(x,-v_0)$ at time $t$ to an arbitrary point at time $t_f$ while staying positive. The combination of the finite velocity of the particle and the meander condition induces a positive single sided light cone in which the particle must remain (shaded red region). Note that once the particle is beyond the line $x=-v_0\,(t-t_f)$ in the $(x,t)$ plane (green dashed line), the particle survives for sure and the tumbling rates return to their free constant value $ \gamma_M^*(x,\dot x,t\,|\,t_f)=\gamma$.}
            \label{fig:meanders}
\end{figure}

A meander is a trajectory that starts at the origin and stays above it, regardless of its final position. The particle must start from the origin $x_0=0$, necessarily in the state $\sigma_0=+1$, and remain above the origin up to time $t_f$:
  \begin{align}
x(0)=0\,,\quad \dot x(0)=+v_0\,,\quad x(t')\geq 0\quad \forall t' \in [0,\,t_f]\,.\label{eq:mea}
  \end{align}
Similarly to the bridge propagator (\ref{eq:Pb}), the propagator for a meander can be written as (see figure \ref{fig:meanders})
\begin{align}
   P_M(x,t,\sigma\,|\,t_f)=\frac{P_{\text{absorbing}}(x,t,\sigma)\,S(x,t_f-t,\sigma)}{S(x=0,t_f,+)}\,,\label{eq:Pm}
\end{align} where the subscript $M$ refers to ``meander'', $P_{\text{absorbing}}$ is the forward propagator in the presence of an absorbing boundary located at the origin defined in the previous section. In expression (\ref{eq:Pm}), $S(x,t,\sigma)$ denotes the survival probability, i.e. the probability that a free particle starting at $x$ in the state $\sigma$ does not cross the origin up to time $t$. 
The survival probability satisfies the same Fokker-Plank equations as the backward propagator (\ref{eq:Q}) but must be solved on the half line with the initial condition $S(x,t=0,\sigma)=\Theta(x)$, where $\Theta$ is the Heaviside step function, i.e. $\Theta(x)=1$ if $x> 0$ and $\Theta(x)=0$ if $x<0$. One can show, again using the differential form (\ref{eq:Qdt}), that the boundary condition must be $S(x=0,t,-)=0$. Following the steps in the previous section, we find that the analog of the effective tumbling rates (\ref{eq:effBQ}) are given by
\begin{subequations}
\begin{align}
   \gamma_M^*(x,\dot x=+v_0,t\,|\,t_f) &= \gamma\, \frac{S(x,\tau,-)}{S(x,\tau,+)}\,,\\[1em]
  \gamma_M^*(x,\dot x=-v_0,t\,|\,t_f) &= \gamma\, \frac{S(x,\tau,+)}{S(x,\tau,-)}\,,
\end{align}
\label{eq:gMS}
\end{subequations}
where $\tau=t_f-t$ and $S$ is the survival probability of the free particle in the presence of an absorbing boundary. Using its expression (recalled in \ref{app:prop}), we find the exact expressions of the transition rates:
  \begin{subequations}
  \begin{align}
    \gamma_M^*(x,\dot x=+v_0,t\,|\,t_f) &=   \gamma\,\frac{1-\int_0^{\tau} dt' F(t',x,-)}{1-\int_0^{\tau} dt' F(t',x,+)}\,,\\
      \gamma_M^*(x,\dot x=-v_0,t\,|\,t_f) &= \gamma\,\frac{1-\int_0^{\tau} dt' F(t',x,+)}{1-\int_0^{\tau} dt' F(t',x,-)}\,, 
  \end{align}
  \label{eq:effM}
  \end{subequations}
where $\tau=t_f-t$. In expression (\ref{eq:effM}), the function $F(t,x,\sigma)$ is the first-passage distribution (see \ref{app:prop}) given by 
  \begin{subequations}
\begin{align}
  F(t,x,+) &= \gamma\,\frac{e^{-\gamma t}}{g(t,x)}\left(\frac{\gamma x}{v_0}\,I_0[\gamma h(t,x)]+\sqrt{\frac{f(t,x)}{g(t,x)}}\,I_1[\gamma h(t,x)]\right)\,,\\
  F(t,x,-) &= \gamma\,e^{-\gamma t}\,\left(\delta[f(t,x)]+\frac{\gamma\, x}{v_0\,\sqrt{h(t,x)}}\,I_1[\gamma h(t,x)]\right)\,,
\end{align}
\label{eq:Fm}
\end{subequations}
where $I_0(z)$ and $I_1(z)$ denote the modified Bessel functions while the functions $f$, $g$, and $h$ are defined in (\ref{eq:fgh}). As in the bridge case, the Dirac delta terms in the effective rates (\ref{eq:effM}) enforce the particle to remain in the positive single sided light cone defined as
 \begin{align}
  0\leq x\leq v_0\, t\,, \quad \text{when }\,\quad 0\leq t \leq t_f\,,\label{eq:pslc}
 \end{align}
which is a natural boundary induced by the combination of the finite velocity of the particle along with the meander constraint. In practice, when performing numerical simulations, these Dirac delta terms can be safely removed from the effective tumbling rates and be replaced by hard constraints such that the particle must remain positive.  Note that once the particle is beyond the line $x=-v_0\,(t-t_f)$ in the $(x,t)$ plane, the particle survives for sure and the tumbling rates return to their free constant value $\gamma$ (see figure \ref{fig:meanders}). 

\begin{figure}[t]
\subfloat{%
 \includegraphics[width=0.5\textwidth]{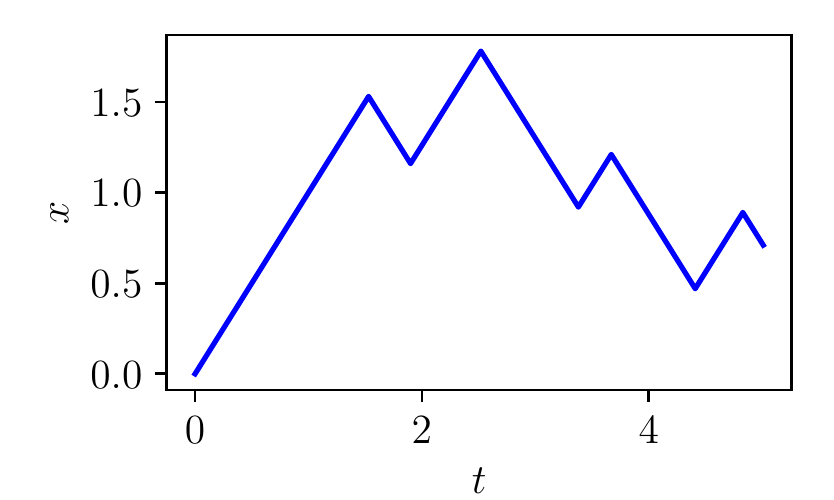}%
}\hfill
\subfloat{%
  \includegraphics[width=0.5\textwidth]{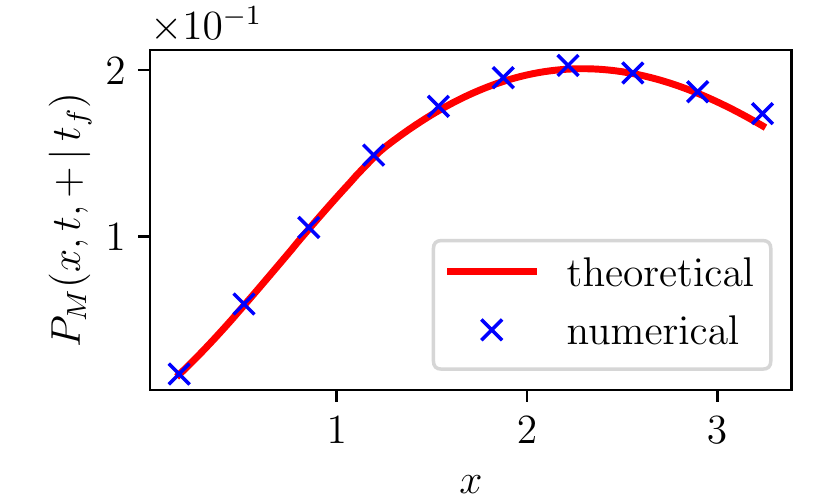}%
}\hfill
\caption{\textbf{Left panel:} A typical meander trajectory of a RTP starting at the origin and remaining positive up to time $t_f=5$. The trajectory was generated using the effective tumbling rates (\ref{eq:effM}). \textbf{Right panel:} Position distribution at $t=3\,t_f/4$ for a meander RTP starting at the origin and remaining positive up to time $t_f=5$. The position distribution $P_M(x,t,+\,|\,t_f)$ obtained numerically by sampling from the effective tumbling rates (\ref{eq:effM}) is compared with the theoretical prediction (\ref{eq:Pmeander}). The agreement is excellent. The distribution exhibits two regimes, one below $x=-v_0(t-t_f)=5/4$ and one beyond $x=5/4$, due to the region in the $(x,t)$ plane where the tumbling rates return to their free constant value $\gamma$ (see figure \ref{fig:meanders}). Note that the Dirac delta function of the never tumbling trajectory is not shown to fit the data within the limited window size.}\label{fig:meander}
\end{figure}

The effective rates (\ref{eq:effM}) generate run-and-tumble meander trajectories (see left panel in figure \ref{fig:meander}).  In the right panel in figure \ref{fig:meander}, we computed numerically the probability distribution of the position at some intermediate time $t=t_f/2$, by generating meander trajectories from the effective tumbling rates (\ref{eq:effM}) and compared it to the theoretical position distribution for the meander propagator which can be easily computed by substituting the free forward  propagator and the survival probability in the presence of an absorbing boundary (recalled in the \ref{app:prop}) in the expression of the meander propagator in (\ref{eq:Pm}):
   \begin{subequations}
\begin{align}
P_M(x,t,+) &= \frac{e^{\gamma\,t_f}}{v_0}\, \frac{F(t,x,-\sigma)\left[1-\int_0^{\tau} dt' F(t',x,\sigma)\right]}{I_0(\gamma\,t_f)+I_1(\gamma\,t_f)}\,,\\
P_M(x,t,-)&=P_M(x,\tau,+)\,,  
\end{align}
\label{eq:Pmeander}
\end{subequations}
where $\tau=t_f-t$ and $F$ is defined in (\ref{eq:Fm}). In the expressions (\ref{eq:Pmeander}), $I_0(z)$ and $I_1(z)$ denote the modified Bessel functions while the functions $f$, $g$, and $h$ are defined in (\ref{eq:fgh}).
As can be seen in figure \ref{fig:meander}, the agreement is excellent.
As in the bridge case, we can compute the diffusive limit (\ref{eq:bml}) of the effective rates (\ref{eq:effM}) to find that they take a rather simple form
\begin{subequations}
\begin{align}
   \gamma_M^*(x,\dot x=+v_0,t\,|\,t_f) &\sim \gamma - \sqrt{\frac{2\gamma}{\pi\tau}}\frac{e^{-\frac{x^2}{4D\tau}}}{\text{erf}\left(\frac{x}{\sqrt{4D\tau}}\right)}+O(\gamma^{-1})\,,\\
   \gamma_M^*(x,\dot x=-v_0,t\,|\,t_f) &\sim\gamma + \sqrt{\frac{2\gamma}{\pi\tau}}\frac{e^{-\frac{x^2}{4D\tau}}}{\text{erf}\left(\frac{x}{\sqrt{4D\tau}}\right)}+O(\gamma^{-1})\,,
\end{align} 
\label{eq:gMbm}
\end{subequations}
which, upon inserting in the effective Fokker-Plank equations (\ref{eq:effFPb}) gives back the effective Langevin equation for Brownian meanders \cite{MajumdarEff15}.

\section{Summary and outlook}
\label{sec:sum}
In this paper, we studied run-and-tumble bridge trajectories, which is a prominent example of a non-Markovian constrained process. We provided an efficient way to generate them numerically by deriving an effective Langevin equation for the constrained dynamics. We showed that the tumbling rate of the RTP acquires a space-time dependency that naturally encodes the bridge constraint. We derived the exact expression of the effective tumbling rate and showed how it yields to an efficient sampling of run-and-tumble bridge trajectories. The method is quite versatile and we extended it to other types of constrained run-and-tumble trajectories such as excursions and meanders.

It would be interesting to generalise our results to higher dimensions and study geometrical properties of bridge trajectories such as their convex hull. Indeed, the convex hull is a natural observable that appears in the study of the motion of foraging animals and measures the spatial extent of their territory \cite{Randon09}. In this context, the bridge constraint would enforce the condition that the animal must return to its home after a fixed amount time. Another possible extension of this work would be to derive effective equations of motion to generate other types of constrained trajectories. For instance, it would be interesting to study various constraints on linear statistics, such as trajectories with a fixed area below the curve.

\section*{Acknowledgments}
  This work was partially supported by the Luxembourg National Research Fund (FNR) (App.  ID 14548297).

\appendix
\section{Useful results on the free run-and-tumble particle}
\label{app:prop}
In this appendix, we recall useful results on the free run-and-tumble particle. A derivation of these results can be found in e.g. \cite{DebruyneSur21}.
\subsection{Free propagator}
The free forward propagator $P(x,t,\sigma|\sigma_0)$ satisfying the Fokker-Plank equation (\ref{eq:P}) on the real line along with the initial condition $P(x,t=0,\sigma|\sigma_0)=\delta_{\sigma,\sigma_0}\,\delta(x)$ is given by
\begin{subequations}
 \begin{align}
    P(x,t,+|+) &=  \left\{\begin{array}{ll}0\, , & v_0 t<|x|\,,\\\frac{\gamma}{v_0}e^{-\gamma t}\left(\delta(v_0 t-x)+\frac{\sqrt{v_0 t+x}}{\sqrt{v_0 t-x}}\,I_1(\gamma\sqrt{t^2-(x/v_0)^2})/2\right)\, ,& v_0 t\geq|x|\,, \\
   \end{array}\right. \\
P(x,t,-|+) &=\left\{\begin{array}{ll} 0\,, & v_0t<|x|\,,\\ \frac{\gamma}{v_0}e^{-\gamma t}\, I_0(\gamma \sqrt{t^2-(x/v_0)^2})/2\,, & v_0 t\geq |x|\,,
\end{array}\right.\\
P(x,t,-|-) &= P(-x,t,+|+)\,,  \\
P(x,t,+|-) &=P(-x,t,-|+) \,,
\end{align}
\label{eq:Papp}
\end{subequations}
where $I_0(z)$ and $I_1(z)$ denote the modified Bessel functions.
The free backward propagator $Q(x,t,\sigma|\sigma_f)$ satisfying the Fokker-Plank equation (\ref{eq:Q}) on the real line along with the initial condition $Q(x,t=0,\sigma|\sigma_f)=\delta_{\sigma,\sigma_f}\,\delta(x)$ is simply given by
\begin{align}
  Q(x,t,\sigma|\sigma_f) = P(x,t,-\sigma_f|-\sigma)\,,\label{eq:Qapp}
\end{align}
where we used the time reversibility of the trajectories in the extended $(x,\dot x)$ space.

\subsection{Survival probability}
The survival probability $S(x,t,\sigma)$, i.e., the probability of a free particle starting at $x$ in the state $\sigma$ does not cross the origin up to time $t$, is given in terms of the first-passage distribution $F(t,x,\sigma)$ as
\begin{align}
  S(x,t,\sigma) = 1 - \int_0^t dt' F(t',x,\sigma)\,,\label{eq:Sapp}
\end{align}
where $F(t,x,\sigma)$ is the probability density that the particle reaches the origin at $t$ given that it started at $x$ in the state $\sigma$:
    \begin{subequations}
\begin{align}
  F(t,x,+) &= \left\{\begin{array}{ll}
  0\,,&t<\frac{x}{v_0}\,,\\
  \gamma\,\frac{e^{-\gamma t}}{g(t,x)}\left(\frac{\gamma x}{v_0}\,I_0[\gamma h(t,x)]+\sqrt{\frac{f(t,x)}{g(t,x)}}\,I_1[\gamma h(t,x)]\right)\,,& t\geq\frac{x}{v_0}\,,
  \end{array}\right.\\[1em]
  F(t,x,-) &=  \left\{\begin{array}{ll}
  0\,,&t<\frac{x}{v_0}\,,\\
  \gamma\,e^{-\gamma t}\,\left(\delta[f(t,x)]+\frac{\gamma\, x}{v_0\,\sqrt{h(t,x)}}\,I_1[\gamma h(t,x)]\right)\,,& t\geq\frac{x}{v_0}\,,
  \end{array}\right.
\end{align}
\label{eq:Fa}
\end{subequations}
where $I_0(z)$ and $I_1(z)$ denote the modified Bessel functions while the functions $f$, $g$, and $h$ are defined in (\ref{eq:fgh}). The integral in the survival probability (\ref{eq:Sapp}) can be evaluated exactly when $x=0$, necessarily when $\sigma=+$, and yields
\begin{align}
  S(x=0,t,+) = e^{-\gamma\,t}\left[I_0(\gamma\,t)+I_1(\gamma\,t) \right]\,.
\end{align}

\subsection{Propagator in the presence of an absorbing boundary}
The forward and backward propagators, $P_{\text{absorbing}}(x,t,\sigma)$ and $Q_{\text{absorbing}}(x,t,\sigma)$, for a particle starting at the origin in the presence of an absorbing boundary satisfy the set of Fokker-Plank equations (\ref{eq:P}) and (\ref{eq:Q}) that must be solved on the half line with the initial condition $P_{\text{absorbing}}(x,t\!=\!0,\sigma)=\delta_{\sigma,+}\delta(x)$ and $Q_{\text{absorbing}}(x,t\!=\!0,\sigma)=\delta_{\sigma,-}\delta(x)$ along with the boundary condition $P_{\text{absorbing}}(x\!=\!0,t,+)=0$ and $Q_{\text{absorbing}}(x\!=\!0,t,-)=0$. The derivation for an arbitrary initial (final) position can be found in \cite{Singh2019}. However, in the simpler case where the particle starts (finishes) at the origin, one can relate the propagators to the first-passage distribution presented in the previous section. Indeed, one can write
\begin{align}
  Q_{\text{absorbing}}(x,t,\sigma) = \frac{1}{v_0}\,F(x,t,\sigma)\,,\label{eq:QAapp}
\end{align}
where $F$ is given in (\ref{eq:Fa}) and the prefactor is the Jacobian of the change of variables $dt/dx=1/v_0$. In addition, we have that
\begin{align}
 P_{\text{absorbing}}(x,t,\sigma) = Q_{\text{absorbing}}(x,t,-\sigma)\,,\label{eq:PAapp}
\end{align}
by using the time reversibility of the trajectories in the extended $(x,\dot x)$ space.

\end{document}